\documentclass[runningheads,a4paper]{llncs}
\usepackage{cite}
\usepackage[utf8]{inputenc}
\usepackage{graphicx}
\usepackage{amsmath}
\usepackage{hyperref}
\usepackage{amsfonts}
\usepackage{array}
\usepackage{extarrows}
\usepackage{caption}
\usepackage{algorithm}
\usepackage{algpseudocode}
\usepackage{setspace}
\algrenewcommand\algorithmicindent{0.90em}%
\newcommand{\CommentX}[1]{\mbox{}\hfill~#1~}
\algblockdefx[IF]{If}{EndIf}[1]{\scriptsize \algorithmicif\ #1\ \algorithmicthen}{\scriptsize\textbf{end}}
\algblockdefx[FOR]{For}{EndFor}[1]{\scriptsize\algorithmicfor\ #1}{\scriptsize\textbf{end}}
\algblockdefx[While]{While}{EndWhile}[1]{\scriptsize\algorithmicwhile\ #1}{\scriptsize\textbf{end}}
\algblockdefx[Procedure]{Procedure}{EndProcedure}[2]{\scriptsize\algorithmicprocedure\ \texttt{#1}(#2)}{\scriptsize\textbf{end}}
\usepackage{amssymb}
\usepackage{float}
\usepackage{multirow}
\usepackage[english]{babel}
\usepackage{appendix}
\usepackage{xparse}
\newcommand*{\QEDB}{\hfill\ensuremath{\square}}
\NewDocumentCommand\Rm{mg}{%
    \IfNoValueTF{#2}
    {\mathbb{R}_{\max}^{#1}}
    {\mathbb{R}_{\max}^{#1 \times #2}}
}
\newcommand{\Rmax}{\mathbb{R}_{\max}}

\usepackage{amsthm}
\theoremstyle{plain}
 \usepackage{tikz}
\usetikzlibrary{arrows,patterns,positioning}

\usepackage{url}
\urldef{\mailsa}\path|{muhammad.syifaul.mufid, alessandro.abate}@cs.ox.ac.uk|    
\urldef{\mailsb}\path| dieky@matematika.its.ac.id|

\begin{document}

\mainmatter  

\title{Bounded Model Checking of Max-Plus Linear Systems via Predicate Abstractions}

\titlerunning{BMC of MPL Systems via PA}

%
%
\author{Muhammad Syifa'ul Mufid$^\dagger$%
\and Dieky Adzkiya$^\ddagger$
\and Alessandro Abate$^\dagger$}
\authorrunning{M. S. Mufid, D. Adzkiya, A. Abate}

\institute{$^\dagger$Department of Computer Science, University of Oxford,United Kingdom\\
\mailsa\vspace*{2ex}\\
$^\ddagger$Department of Mathematics, Institut Teknologi Sepuluh Nopember\\
 Surabaya, Indonesia\\
\mailsb\\
}

%
%

\maketitle

\begin{abstract}
This paper introduces the abstraction of max-plus linear (MPL) systems via predicates. 
Predicates are automatically selected from system matrix, 
as well as from the specifications under consideration. 
We focus on verifying time-difference specifications, 
which encompass the relation between successive events in MPL systems.  
We implement a bounded model checking (BMC) procedure over a predicate abstraction of the given MPL system, 
to verify the satisfaction of time-difference specifications. 
Our predicate abstractions are experimentally shown to improve on existing MPL abstractions algorithms. 
Furthermore, 
with focus on the BMC algorithm, 
we can provide an explicit upper bound on the completeness threshold by means of the transient and the cyclicity of the underlying MPL system.
%
\end{abstract}

\section{Introduction}

Max-Plus-Linear (MPL) systems are a class of discrete-event systems, 
with dynamics based on two binary operations (maximisation and addition) over a max-plus semiring. 
MPL systems are used to model synchronisation phenomena without concurrency. 
These systems have been used in many areas, 
such as manufacturing \cite{Aleksey}, transportation \cite{Heidergott}, and biological systems \cite{Chris,Comet}. 

Classical analysis of MPL systems is conducted using algebraic approaches \cite{Baccelli,Heidergott}. 
Recently, an alternative take based on formal abstractions has been developed to verify MPL systems against quantitative specifications \cite{Dieky1} that are general and expressive. 
The performance and scalability of the abstraction approach has been later improved by employing tropical operations \cite{Mufid} that are native to the max-plus semiring.  

This work pushes the envelop on scalability of formal abstractions of MPL systems. 
We newly apply predicate abstractions (PA) and bounded model checking (BMC) for the verification of MPL systems over time-difference specifications. 
Predicate abstractions are an abstraction approach that leverage a set of predicates, 
and have been classically used for software and hardware verification \cite{Flanagan,Clarke3}, 
for the abstraction of programs \cite{Clarke5,Ball}, 
and for reachability analysis of hybrid systems \cite{Alur2}.  

BMC is a symbolic model checking approach that leverages SAT solvers. 
The basic idea is to attempt finding counterexamples with a length bounded by some integer. 
If no counterexamples are found, the length is greedily increased. 
The approach is sound (counterexamples are correct), and complete (no counterexamples are admitted) whenever a completeness threshold (CT) for the length is reached \cite{Biere1,Biere2}. 
Whilst there exist results on correct upper-bounds on the CT, in practice BMC is run until  the underlying problem becomes intractable. 


This paper has two specific contributions.  
The first contribution is related to the abstraction approach.
Moving beyond \cite{Dieky1,Mufid}, 
where the abstraction procedures are based on the translation of MPL systems into piecewise affine (PWA) systems, 
in this work we newly employ PA.  
Namely, we determine a set of predicates such that the dynamics within each partitioning region is affine. 
In other words, there is no need to compute PWA systems anymore. 

The second contribution is related to the model-checking approach. 
\cite{Dieky1} employs standard model checking to verify the abstract transition system. 
In this paper, we leverage BMC: 
notice that PA naturally yield Boolean encodings that can be relevant for the SAT-based BMC procedure. 
We focus on time-difference specifications. 
Since we are working on abstractions, 
counterexample generated by the BMC procedure needs to be checked for spuriousness (cf. \autoref{alg5} and \autoref{alg6}). 
Whenever a counterexample is spurious, 
we refine the abstract transition using the procedure in \cite{Dieky3}, 
combined with lazy abstraction \cite{Henzinger}. 
Finally, for the considered time-difference specifications, 
we show that the CT can be upper-bounded by means of the transient and cyclicity of the concrete MPL system - such bounds are in general tighter than those obtained working on the abstract transition system. 
As a side result, we provide a few instance of ``direct verification'',  
where the model checking of MPL models can be performed straightforwardly for time-difference specifications. 

The paper is organised as follows. Section 2 describes the basics of models, abstraction techniques and temporal logic formulae used in this work.  
It also contains the notion of time-difference over MPL systems. 
The contributions of this paper are contained in Sections 3 and 4. 
%
%
The comparison of abstraction procedures is presented in Section 5, 
with PA implemented in C++ and model checking run over NuSMV \cite{nusmv}.  
We also compare the completeness threshold w.r.t. transient and cyclicity of MPL systems with those that are computed by NuSMV. The paper is concluded in Section 6.

\section{Model and Preliminaries}
\subsection{Max-Plus Linear Systems}
By max-plus semiring we understand an algebraic structure $(\Rmax ,\oplus,\otimes)$ where $
\Rmax:=\mathbb{R}\cup\{\varepsilon:= - \infty\}$ and $a\oplus b:=\max\{a,b\},~a\otimes b:=a+b~~~~\forall a,b\in \Rmax.$ The set of $n\times m $ matrices over max-plus semiring is denoted as $\Rm{n}{m}$. Two binary operations of a max-plus semiring can be extended to matrices as follows
\begin{align*}
 [A\oplus B](i,j) &= A(i,j)\oplus B(i,j),\\
 [A\otimes C](i,j)&= \bigoplus_{k=1}^m A(i,k)\otimes C(k,j),
\end{align*}
where $A,B\in \Rm{n}{m}, C\in \Rm{m}{p}$. Given $r\in\mathbb{N}$, the max-plus algebraic power of $A\in \Rm{n}{n}$ is denoted by $A^{\otimes r}$ and corresponds to $A\otimes \ldots \otimes A$ ($r$ times). 

A Max-Plus Linear (MPL) system is defined as
\begin{equation}
\textbf{x}(k+1) = A\otimes \textbf{x}(k),
\label{mpl}
\end{equation}
where $A\in \Rm{n}{n}$ is the system matrix and $\textbf{x}(k) = [x_1(k)\ldots x_n(k)]^\top$ is the state variables \cite{Baccelli}. 
In particular, for $i\in \{1,\ldots,n\}$, $x_i(k+1)=\max\{A(i,1)+x_1(k),\ldots,A(n,i)+x_n(k)\}$. 
In applications, $\textbf{x}$ represents the time stamps of the discrete events, while $k$ corresponds to the event counter. 
Therefore, it is more convenient to take $\mathbb{R}^n$ (instead of $\Rm{n}$) as the state space.

\begin{definition}[Precedence Graph {\cite{Baccelli}}]
\normalfont
The precedence graph of $A$, denoted by $\mathcal{G}(A)$, is a weighted directed graph with nodes $1,\ldots,n$ and an edge from $j$ to $i$ with weight $A(i,j)$ if $A(i,j)\neq \varepsilon$. \QEDB
\end{definition}
\begin{definition}[Regular Matrix \cite{Heidergott}] 
\normalfont
A matrix $A\in \mathbb{R}_{\max}^{n\times n}$ is called regular if there is at least one finite element in each row. 
\QEDB
\end{definition}
\begin{definition}[Irreducible Matrix \cite{Baccelli}] 
\normalfont
A matrix $A\in \mathbb{R}_{\max}^{n\times n}$ is called irreducible if the corresponding precedence graph $\mathcal{G}(A)$ is strongly connected. \QEDB
\end{definition}

Recall that a directed graph is strongly connected if for two different nodes $i,j$ of the graph, there exists a path from $i$ to $j$ \cite{Baccelli,Bart2}. The weight of a path $p=i_1i_2\ldots i_k$ is equal to the total weight of the corresponding edges i.e. $|p|=A(i_2,i_1)+\ldots+A(i_{k},i_{k-1})$. A circuit, namely a path that begins and ends at the same node, is called \textit{critical} if it has maximum average weight, which is the weight divided by the length of path \cite{Baccelli}.

Every irreducible matrix $A\in\Rmax^{n\times n}$ admits a unique  max-plus eigenvalue $\lambda\in\mathbb{R}$, which corresponds to the weight of critical circuit in $\mathcal{G}(A)$. Furthermore, by \autoref{trans_con} next, $A$ satisfies the so-called transient condition:  
\begin{proposition}[Transient Condition\cite{Baccelli}] 
\label{trans_con}
\normalfont
For an irreducible matrix $A\in\Rmax^{n\times n}$ and its corresponding max-plus eigenvalue $\lambda\in\mathbb{R}$, there exist $k_0,c\in\mathbb{N}$ such that $A^{\otimes(k+c)}=\lambda c\otimes A^{\otimes k}$ for all $k\geq k_0$. The smallest such $k_0$ and $c$ are called the transient and the cyclicity of $A$, respectively. \QEDB
\end{proposition}

\begin{example}
Consider a $2\times 2$ MPL system that represents a simple railway network \cite{Heidergott}: 
\begin{equation}
\label{mpl_ex}
\textbf{x}(k+1) = \begin{bmatrix}
2 &~5 \\ 3 &~3
\end{bmatrix} \otimes \textbf{x}(k).
\end{equation}
Its max-plus eigenvalue is $\lambda=4$, whereas the transient and cyclity for the matrix are $k_0=c=2$. \QEDB
\end{example}

Any given MPL system can be translated into a Piece-Wise Affine (PWA) system \cite{Heemels}. 
A PWA system comprises of spatial regions with corresponding PWA dynamics. 
The regions are generated from all possible coefficients $\textbf{g}=(g_1,\ldots,g_n)\in \{1,\ldots,n\}^n$, 
which 
satisfies
$A(i,g_i)\neq \varepsilon $ for $ 1\leq i\leq n$. As shown in \cite{Dieky1}, the region corresponding to $\textbf{g}$ is
\begin{equation}
\mathsf{R}_\textbf{g} = \bigcap_{i=1}^n\bigcap_{j=1}^n\left\{\textbf{\text{x}}\in \mathbb{R}^n|x_{g_i}-x_j\geq A(i,j)-A(i,g_i)\right\}.
\label{pwa}
\end{equation}
One could check that for each non-empty $\mathsf{R}_\textbf{g}$ and $\textbf{x}(k)\in \mathsf{R}_\textbf{g}$, the MPL system \eqref{mpl} can be rewritten as the following affine dynamics: 
\begin{equation}
 x_i(k+1)=x_{g_i}(k)+A(i,g_i),~~ i=1,\ldots,n.
\label{af}
\end{equation}
Notice that \eqref{af} can be expressed as $\textbf{x}(k+1)=A_\textbf{g}\otimes \textbf{x}(k)$, where $A_\textbf{g}$ is a region matrix \cite{Mufid} for the coefficient $\textbf{g}$.

\subsection{Time Differences in MPL Systems}

We consider delays occurring between events governed by \eqref{mpl}. 
Delays can describe the difference of two events corresponding to the same event counter but at different variable indices (i.e. $x_i(k)-x_j(k)$), 
or the difference of two consecutive events for the same index (i.e. $x_i(k+1)-x_i(k)$). This paper focuses on the later case although, in general, the results of this paper can be applied to the former case.

We write the $(k+1)^\text{th}$ time difference for the $i^\text{th}$ component as $t_i(k)=x_i(k+1)-x_i(k)$. 
One can see that 
\begin{eqnarray}
\label{td}
t_i(k)&=& \max_{{j^\ast}\in \texttt{fin}_i}\{x_{{j^\ast}}(k)+A(i,{{j^\ast}})\} -x_i(k),
\end{eqnarray}
where $\texttt{fin}_i$ is the set containing the indices of finite elements of $A(i,\cdot)$.\footnote{For the sake of simplicity, we write the elements of $\texttt{fin}_i$ in a strictly increasing order.}

\subsection{Transition Systems and Linear Temporal Logic}

\begin{definition}[Transition System \cite{Baier}]
\normalfont
A transition system is formulated by a tuple $(S,T,I,\mathcal{AP},L)$, where
\begin{itemize}
\item[$\bullet$] $S$ is a set of states,
\item[$\bullet$] $T\subseteq S\times S$ is a transition relation,
\item[$\bullet$] $I\subseteq S$ is a set of initial states,
\item[$\bullet$] $\mathcal{AP}$ is a set of atomic propositions, and
\item[$\bullet$] $L:S\rightarrow 2^{\mathcal{AP}}$ is a labelling function. \QEDB
\end{itemize}
\end{definition}

%

A \textit{path} of $TS$ is defined as a sequence of states $\pi=s_0s_1\ldots$, where $s_0\in I$ and $(s_i,s_{i+1})\in T$ for all $i\geq 0$. 
We denote $\pi[i]=s_{i-1}$ as the $i^\text{th}$ state of $\pi$. Furthermore, $|\pi|$ represents the number of transitions in $\pi$. 


Linear temporal logic (LTL) is one of the predominant logics that are used for specifying properties over the set of atomic propositions \cite{Baier}. 
LTL formulae are recursively defined as follows. 
\begin{definition}[Syntax of LTL \cite{Baier}]
\normalfont
LTL formulae over the set of atomic propositions $\mathcal{AP}$ are constructed according to the following grammar:
\[\varphi:=\text{true}~|~a~|~\varphi_1 \wedge \varphi_2~|~\neg \varphi~|~\bigcirc \varphi~|~\varphi_1 ~\mathsf{U}~\varphi_2,\]
where $a\in \mathcal{AP}$.
\QEDB
\end{definition}

The symbol $\bigcirc$ (next) and $\mathsf{U}$ (until) are called temporal operators. Two additional operators, $\lozenge $ (eventually) and $\square$ (always), are generated via the until operators: 
$\lozenge \varphi = \text{true}~\mathsf{U}~\varphi~~\text{and}~~\square \varphi=\neg \lozenge \neg \varphi$. We refer to \cite{Baier} for the semantics of LTL formulae including the satisfaction relation $\models$ over transition systems.


\subsection{Abstractions and Predicate Abstractions}

Abstractions are techniques to generate a finite and smaller model from a large or even infinite-space (i.e., a continuous-space model, e.g., an MPL system) model. 
Abstractions can reduce the verification of a temporal property $\varphi$ over the original model (a \textit{concrete} model with state space $S$), to checking a related property on a simpler \textit{abstract} model (over $\hat{S}$) \cite{Baier}. The mapping from $S$ to $\hat{S}$ is called \textit{abstraction function}.

From a (concrete) transition system $TS=(S,T,I,\mathcal{AP},L)$ and an abstraction function $f:S\rightarrow \hat{S}$, the (abstract) transition system $TS_f = (\hat{S},T_f,I_f,\mathcal{AP},L_f)$ is generated from $TS$ as follows: i) $I_f=\{f(s)\mid s\in I\}$, ii) $(f(s)$,$f(s^\prime))\in T_f$ if $(s,s^\prime)\in T$, and iii) $L_f(f(s))=L(s)$, for all $s\in S$.

The important relation between $TS$ and $TS_f$ is that the former is \textit{simulated} by the latter (which is denoted by $TS \preceq TS_f)$. In detail, all behaviour on concrete transition system occur on the abstract one. The formal definition of simulation relation can be found in \cite[Definition 7.47]{Baier}. Furthermore, given an LTL formula $\varphi$, $TS_f\models \varphi$ implies $TS\models \varphi$ \cite{Baier,Clarke2}.

Predicate abstractions \cite{Graf,Clarke1,Clarke2,Das} denote abstraction methods that use a set of \textit{predicates} 
$P=\{p_1,\ldots,p_k\}$ to characterise the abstract states.  
Predicates are identified from the concrete model, and possibly from the specification(s) under consideration. 
Each predicate $p_i$ corresponds to a Boolean variable $b_i$ and each abstract state $\hat{s}\in \hat{S}$ corresponds to a Boolean assignment of these $k$ Boolean variables \cite{Clarke2}. Therefore, we obtain that $|\hat{S}|\leq 2^k$. An abstract state will be labelled with predicate $p_i$ if the corresponding $b_i$ is $\text{true}$ in that state. For this reason, predicates also serve as atomic propositions \cite{Clarke2}.

The predicates are also used to define an abstraction function between the concrete and abstract state spaces. A concrete state $s\in S$ will be related to an abstract state $\hat{s}\in \hat{S}$ iff the truth value of $p_i$ on $s$ equals the value of $b_i$ on $\hat{s}$. The abstraction function for predicate abstractions is defined as
$f(s) = \bigwedge_{i=1}^k \text{val}(s,p_i)$,
where $\text{val}(s,p_i)=b_i$ if $p_i$ is satisfied in $s$, otherwise $\neg b_i$.


\section{Predicate Abstractions of MPL Systems}
\subsection{Related Work}
The notion of abstractions of an MPL system has been first introduced in \cite{Dieky1}: 
there, it leverages translation of an MPL system into the corresponding PWA system. 
The resulting abstract states are expressed as Difference-Bound Matrices (DBM). 
A more efficient procedure for MPL abstractions via max-plus algebraic operations is later discussed in \cite{Mufid}. 

\subsection{Generation of the Predicates}
Considering an abstraction via a set of predicates, the first issue is to find appropriate predicates. 
Recall that related abstraction techniques \cite{Dieky1 ,Mufid} explore the connection between MPL and PWA systems and use DBMs to represent the abstract states. 
Similarly, predicates here are chosen such that the dynamics in the resulting abstract states are affine as in \eqref{pwa} and can be expressed as DBMs. 
Following these considerations, the predicates are defined as an inequality $p\equiv x_i-x_j \sim c$ where $\sim \hspace*{1ex}\in\{>,\geq\}\footnote{In this paper, we always use $p\equiv x_i-x_j\geq c$ as a predicate.},c\in \mathbb{R}$. For simplicity, we may write a predicate as a tuple $p\equiv (i,j,c,s)$ where $s=1$ if $\sim ~=~\geq$, otherwise $s=0$. The negation of $p$ then can be written as $\neg p\equiv (j,i,-c,1-s)$.

From the PWA region in \eqref{pwa}, $c$ can be chosen from the difference of two finite elements of the state matrix $A\in \Rm{n}{n}$ at the same row. In detail, if $A(k,j)\neq \varepsilon$ and $A(k,i)\neq \varepsilon$ with $i<j$ and $1\leq k\leq n$, then we get a predicate $(i,j,A(k,j)-A(k,i),1)$. 

\autoref{alg1} shows a procedure to generate the predicates from an MPL system. For each $k\in \{1,\ldots,n\}$, $P_k$ is a set of predicates generated from $A(k,\cdot)$. If there are exactly $m>1$ finite elements at each row of $A$ then $|P_k|=\binom{m}{2}$ and in general $|\bigcup _{k=1}^n P_{k}|\leq n\binom{m}{2}$: indeed, it is possible to get the same predicate from two different rows when $A(k_1,j)-A(k_1,i)=A(k_2,j)-A(k_2,i)$ for $k_1\neq k_2$. 
\begin{algorithm}[!ht]
 \scriptsize
\hspace*{\algorithmicindent} \textbf{Input:} $A\in \Rm{n}{n}$,   \\
\hspace*{\algorithmicindent} \textbf{Output:} $P_{mat}$, a set of predicates 
\caption{\small Generation of predicates from an MPL system}\label{alg1}
\begin{algorithmic}[1]
\Procedure{mpl2pred}{$A,k$} \Comment{\scriptsize generation of predicates from the $k^\text{th}$ row of $A$}
\State $P_{k}\gets \emptyset$
\State $\texttt{fin}_k:=\texttt{Find}(A(k,\cdot)\neq \varepsilon)$ \Comment{\scriptsize $\texttt{fin}_k$ is a vector consisting the index of }
\For{$j\in \{2,\ldots,|\texttt{fin}_k|\}$}\CommentX{\scriptsize finite elements of $A(k,\cdot)$, $\texttt{fin}_k[i]$ is \hspace*{4ex}}
\For{$i\in \{1,\ldots,j-1\}$} \CommentX{\scriptsize the $i^\text{th}$ element of $\texttt{fin}_k$ \hspace*{17.5ex}}
\State $P_{k}\gets P_{k}\cup \{(\texttt{fin}_k[i],\texttt{fin}_k[j],A(k,\texttt{fin}_k[j])-A(k,\texttt{fin}[i]),1)\}$ 
\EndFor
\EndFor        
\State \textbf{return} $P_{k}$
\EndProcedure
\Statex
\Procedure{mpl2pred}{$A$} \Comment{\scriptsize generation of predicates from matrix $A$ }
\State $P_{mat}\gets \emptyset$ 
\For{$k\in\{1,\ldots,n\}$} \Comment{\scriptsize generation of predicates for each row of matrix $A$ }
\State $P_{mat}\gets P_{mat}\cup \texttt{mpl2pred}(A,k)$ \CommentX{\scriptsize and storing the resulting predicates in $P_{mat}$ \hspace*{5ex}}
\EndFor
\State \textbf{return} $P_{mat}$
\EndProcedure
\end{algorithmic}
\end{algorithm}


As mentioned before, predicates can also be associated to given specifications. 
In this paper, we focus on time-difference specifications that are generated from a set of \textit{time-difference propositions}.  
For $\alpha\in \mathbb{R}$, we define a time-difference proposition `$t_i\sim \alpha$' to reason the condition that $x_i^\prime-x_i\sim \alpha$. We remove the counter event $k$ for the sake of simplicity. 


One can rewrite \eqref{td} as $t_i= \max_{{j^\ast}\in \texttt{fin}_i}\{x_{{j^\ast}}+A(i,{{j^\ast}})\} -x_i $. Therefore, from $t_i \sim \alpha $ for $\sim \hspace*{1ex}\in\{>,\geq,<,\leq\}$ we have $\max_{{j^\ast}\in \texttt{fin}_i}\{x_{{j^\ast}}+A(i,{{j^\ast}})\} -x_i \sim \alpha $. 
The number of predicates corresponding to `$t_i\sim \alpha$' is bounded by $|\texttt{fin}_i|$. For each $j^\ast\in \texttt{fin}_i$ we get a predicate $x_{j^\ast}-x_i\sim \alpha-A(i,j^\ast)$. However, in case of $i\in \texttt{fin}_i$, or in other words $A(i,i)\neq \varepsilon$, $x_{i}-x_i\sim \alpha-A(i,j^\ast)$ is not a predicate. \autoref{alg2} shows how to generate the predicates w.r.t. a time-difference proposition. 
\begin{algorithm}[!ht]
 \scriptsize
\hspace*{\algorithmicindent} \textbf{Input:} $A\in \Rm{n}{n}$, a matrix containing exactly $m$ finite elements in each row\\ 
\hspace*{\algorithmicindent} ~~~~~~~~~ $t_i\sim \alpha$, a time-difference proposition\\
\hspace*{\algorithmicindent} \textbf{Output:} $P_{time}$, a set of predicates
\caption{\small Generation of predicates from a time-difference proposition}\label{alg2}
\begin{algorithmic}[1]
\Procedure{td2pred}{$A,t_i\sim \alpha$} 
\State $P_{time}\gets \emptyset$
\State $A(i,i)\gets \varepsilon$
\State $\texttt{fin}_i\gets \texttt{Find}(A(i,\cdot)\neq \varepsilon)$ 
\If{$\sim~ \in \{>,\geq \}$}
\For {$j^\ast \in \texttt{fin}_i$}
\State $P_{time}\gets P_{time}\cup \{(j^\ast,i,\alpha-A(i,j^\ast),s)\}$ \Comment{\scriptsize $s$ is 0 if $\sim$ is $>$ and $s$ is 1 if $\sim$ is $\geq$ }
\EndFor
\ElsIf{$\sim~ \in \{<,\leq \}$}
\For{$j^\ast \in \texttt{fin}_i$}
\State $P_{time}\gets P_{time}\cup \{(i,j^\ast,A(i,j^\ast)-\alpha,s)\}$ \Comment{\scriptsize each predicate uses operator $>$ or $\geq$}
\EndFor
\EndIf 
\State \textbf{return} $P_{time}$
\EndProcedure
\end{algorithmic}
\end{algorithm}

\subsection{Generation of Abstract States} 
This section starts by describing the procedure to generate abstract states via a set of predicates. We denote $P$ as the set of predicates generated by \autoref{alg1} and \autoref{alg2}, i.e. $P=P_{mat}\cup P_{time}=\{p_1,\ldots,p_k\}$. Let $\hat{S}$ be a set of abstract states defined over Boolean variables $B=\{b_1,\ldots,b_k\}$, where the truth value of $b_i$ depends on that of $p_i$. For each Boolean variable $b_i$, we define the corresponding DBM as follows: 
$\texttt{DBM}(b_i)=\{\textbf{x}\in \mathbb{R}^n\mid  p_i~\text{is true in}~ \textbf{x}\}$ and $\texttt{DBM}(\neg b_i)=\{\textbf{x}\in \mathbb{R}^n\mid  p_i~\text{is false in}~ \textbf{x}\}$.
One can show that $\texttt{DBM}(b_i \wedge b_j)=\texttt{DBM}(b_i)\cap \texttt{DBM}(b_j)$. 

\autoref{alg3} shows the steps to generate the abstract states of an MPL system given a set of predicates $P$. 
For each $i\in \{1,\ldots,|P|\}$, we manipulate DBMs: the complexity of \autoref{alg3} depends on emptiness checking of DBM (line 11), which runs in $\mathcal{O}(n^3)$, where $n$ is the dimension of the state matrix \cite{Dieky1}. Therefore, the worst-case complexity of \autoref{alg3} is $\mathcal{O}(2^{|P|}n^3)$.
\begin{algorithm}[!ht]
 \scriptsize
\hspace*{\algorithmicindent} \textbf{Input:} $P$, a set of predicates \Comment{\scriptsize $P=P_{mat}\cup P_{time}$}\\
\hspace*{\algorithmicindent} \textbf{Output:} $\hat{S}$, a set of abstract states\\
\hspace*{\algorithmicindent} ~~~~~~~~~~~ $D$, a partition of $\mathbb{R}^n$ w.r.t. $\hat{S}$\Comment{\scriptsize $D$ is a set of DBMs}
\caption{\small Generation of the abstract states from a set of predicates}\label{alg3}
\begin{algorithmic}[1]
\Procedure{pred$\_$abs}{$P$} 
\State $B\gets \{b_1,\ldots,b_{|P|}\}$ \Comment{\scriptsize a set of Boolean variables}
\State $D\gets \{\mathbb{R}^n\}$
\State $\hat{S}\gets \{\texttt{true}\}$
        \For{$i \in \{1,\ldots,|P|\}$}
        \State $\hat{S} \gets \bigcup_{\hat{s} \in \hat{S}} \{ \hat{s} \wedge \neg b_i \} \cup \bigcup_{\hat{s} \in \hat{S}} \{ \hat{s} \wedge b_i \}  $ 
        \State $D_{neg}\gets \bigcup_{E\in D} \{ E\cap DBM(\neg b_i) \}$ \Comment{\scriptsize each DBM in $D$ is intersected with $DBM(\neg b_i)$ }
        \State $D_{pos}\gets \bigcup_{E\in D} \{ E\cap DBM(b_i) \}$ \Comment{\scriptsize both $D_{neg}$ and $D_{pos}$ are set of DBMs}
        \State $D \gets D_{neg} \cup D_{pos}$ 
        \State $D_{temp}\gets \emptyset $ \Comment{\scriptsize temporary variable for $D$}
        \State $ \hat{S}_{temp}\gets \emptyset$ \Comment{\scriptsize temporary variable for $\hat{S}$}
        \For{$j\in \{1,\ldots,|D|\}$}
        \If{$D[j]$ is not empty} \Comment{\scriptsize DBM emptiness check}
        \State \textbf{add} $D[j]$ to $D_{temp}$ 
        \State \textbf{add} $\hat{S}[j]$ to $\hat{S}_{temp}$ 
        \EndIf
        \EndFor
        \State $D\gets D_{temp}$
        \State $\hat{S}\gets \hat{S}_{temp}$
        \EndFor        
\State \textbf{return} $(\hat{S},D)$
\EndProcedure
\end{algorithmic}
\end{algorithm}
\subsection{Generation of Abstract Transitions} 

Having obtained the abstract states, one needs to generate the abstract transitions, which can be obtained via one-step reachability, as described in \cite{Dieky1}. 
Namely, there is a transition from $\hat{s}_i$ to $\hat{s}_j$ if $\mathsf{Im}(\texttt{DBM}(\hat{s}_i))\cap \texttt{DBM}(\hat{s}_j)\neq \emptyset$, where $\texttt{Im}(\texttt{DBM}(\hat{s}_i))=\{A\otimes \textbf{x}\mid\textbf{x}\in \texttt{DBM}(\hat{s}_i)\}$. The computation of $\texttt{Im}(\texttt{DBM}(\hat{s}_i))$ corresponds to the image of $\texttt{DBM}(\hat{s}_i)$ w.r.t. the affine dynamics of $\hat{s}_i$ which has complexity $\mathcal{O}(n^2)$ \cite{Mufid}.

However, unlike \cite[Algorithm 2]{Mufid}, \autoref{alg3} does not produce the affine dynamics for each abstract state. For each $\hat{s}\in \hat{S}$, we need to find $\textbf{g}$ as in \eqref{af}. One can generate the affine dynamics for $\hat{s}\in \hat{S}$ from the value (either true or false) of $p\in P_{mat}$ on $\hat{s}$. Given a predicate $p\equiv (i,j,c,s)$, we call $i$ and $j$ as the left and right index of $p$ (as $x_i\sim x_j+c$) and denoted them by $\texttt{left}(p)$ and $\texttt{right}(p)$, respectively. 

If $p\equiv (i,j,A(k,j)-A(k,i),1)$ is true in $\hat{s}$, we have $x_i+A(k,i)\geq x_j + A(k,j)$, otherwise $x_j+A(k,j)> x_i + A(k,i)$.  
Hence, the left index of predicates can be used to determine the affine dynamics. 
\autoref{alg4} provides the procedure to find the affine dynamic associated to $\hat{s}\in \hat{S}$.  
\begin{algorithm}[!ht]
 \scriptsize
 \hspace*{\algorithmicindent} \textbf{Input:} $A\in \Rm{n}{n}$, a $m$-regular matrix with $m>1$ \\
\hspace*{\algorithmicindent} ~~~~~~~~~ $\hat{s}\in \hat{S}$, an abstract state \\
\hspace*{\algorithmicindent} ~~~~~~~~~ $P_1,\ldots,P_n$, sets of predicates generated by \autoref{alg1} \\
\hspace*{\algorithmicindent} \textbf{Output:} $\textbf{g}$, the finite coefficient representing the affine dynamics for $\hat{s}$  
\caption{\small Generation of the affine dynamics for an abstract state}\label{alg4} 
\begin{algorithmic}[1]
\Procedure{get$\_$affine}{$A,\hat{s},P_1,\ldots,P_n$} 
\State $\textbf{g}\gets \texttt{zeros}(1,n)$
\For{$k\in \{1,\ldots,n\}$}
\State $\texttt{fin}_k\gets \texttt{Find}(A(k,\cdot)\neq \varepsilon)$ \Comment{\scriptsize recall that elements in $\texttt{fin}_k$ is }
\For{$p\in P_k$}  \CommentX{\scriptsize  in strictly-increasing order \hspace*{3.2ex}}
\If{$p$ is false in $\hat{s}$} 
\State  \textbf{{swap}} $\texttt{left}(p)$ with $\texttt{right}(p)$ in $\texttt{fin}_k$ 
\EndIf
\EndFor
\State $\textbf{g}[k]\gets\texttt{fin}_k[1]$ \Comment{\scriptsize insertion of the $k^\text{th}$ element of $\textbf{g}$}
\EndFor
\State \textbf{return} $\textbf{g}$
\EndProcedure
\end{algorithmic}
\end{algorithm}

For each $k$, $\texttt{fin}_k$ is computed. Initially, the elements of $\texttt{fin}_k$ are in strictly increasing order. Then, for each predicate $p\in P_k$, we swap the location of $\texttt{left}(p)$ and $\texttt{right}(p)$ whenever $p$ is false on $\hat{s}$. Suppose $i$ is the first element of $\texttt{fin}_k$ after swapping. One could show that $x_{i}+A(k,i)\sim x_j+A(k,j)$
for all $j\in \texttt{fin}_k\setminus\{i\}$.

\subsection{Model Checking MPL Systems over Time-Difference Specifications: Direct Verification} 

This section discusses the verification of MPL systems over time-difference specifications. First, we define a (concrete) transition system w.r.t. a given MPL system.
\begin{definition}[Transition system associated with MPL system]
\label{ts_mpl}\newline
\normalfont
A transition system $TS$ for an MPL system in \eqref{mpl} is a tuple $(S,T,\mathcal{X},\mathcal{AP},L)$ where
\begin{itemize}
\item[$\bullet$] the set of states $S$ is $\mathbb{R}^n$,
\item[$\bullet$] $(\textbf{x},\textbf{x}^\prime)\in T$ if $\textbf{x}^\prime=A\otimes \textbf{x}$,
\item[$\bullet$] $\mathcal{X}\subseteq \mathbb{R}^n$ is a set of initial conditions,
\item[$\bullet$] $\mathcal{AP}$ is a set of time-difference propositions,
\item[$\bullet$] the labelling function $L:S\rightarrow 2^{\mathcal{AP}}$ is defined as follows: a state $\textbf{x} \in S$ is labeled by `$t_i\sim \alpha$' if $[A\otimes \textbf{x}-\textbf{x}]_i\sim \alpha$, where $\sim~\in \{>,\geq,<,\leq\}$.\QEDB
\end{itemize}
\end{definition}
We express the time-difference specifications as LTL formulae over a set of time-difference propositions.\footnote{Notice that, in \autoref{ts_mpl} we consider $\mathcal{AP}$ as a set of time-difference propositions.} For instance, $\bigcirc (t_i\leq \alpha)$ represents `the next time difference for the $i^\text{th}$ component is $\leq \alpha$' while $\lozenge \square (t_i\leq \alpha)$ corresponds to `after some finite executions, the time difference for the $i^\text{th}$ component is always $\leq \alpha$'. To check the satisfaction of these specifications, we generate the abstract version of MPL system. 

The abstract transition system $TS_f=(\hat{S},T_f,I_f,P_{mat}\cup P_{time},L_f)$ for an MPL system is generated via predicate abstraction where $P_{mat}$ and $P_{time}$ is the set of predicates generated by \autoref{alg1} and \autoref{alg2}, respectively.  The (abstract) labelling function $L_f$ is defined over predicates $p\in P_{mat}\cup P_{time}$: for $\hat{s}\in \hat{S}$, $p\in L_f(\hat{s})$ iff $p$ is true in $\hat{s}$.
We show the relation between predicates in $P_{time}$ and a time-difference proposition in $\mathcal{AP}$.
\begin{proposition}
\normalfont
Suppose $P_{time}$ is a set of predicates corresponding to a time-difference proposition `$t_i\sim \alpha$' and an abstract state $\hat{s}\in \hat{S}$. 
\begin{itemize}
\item[i.] For $\sim \{>,\geq\}$. A (concrete) state $\textbf{x}\in \texttt{DBM}(\hat{s})$ is labeled by `$t_i\sim \alpha$' iff at least one predicate in $P_{time}$ is true in $\hat{s}$.
\item[ii.] For $\sim \{<,\leq\}$. A (concrete) state $\textbf{x}\in \texttt{DBM}(\hat{s})$ is labeled by `$t_i\sim \alpha$' iff all predicates in $P_{time}$ are true in $\hat{s}$. 
\end{itemize}
\textbf{Proof.} We only need to show the proof for $\sim ~=~ \geq$ and $\sim ~=~ \leq$.
\begin{itemize}
\item[i.] Notice that $[A\otimes \textbf{x}-\textbf{x}]_i\geq \alpha$ is equivalent to $\max_{{j^\ast}\in \texttt{fin}_i}\{x_{{j^\ast}}+A(i,{{j^\ast}})\} \geq x_i+\alpha.$ 
This inequality is satisfied iff at least one of $x_{{j^\ast}}+A(i,{{j^\ast}})\geq x_i+\alpha$ for $j^\ast\in\texttt{fin}_i$ is true. It is indeed equivalent to a predicate $x_{{j^\ast}}-x_i \geq \alpha - A(i,{{j^\ast}})\in P_{time}$.
\item[ii.] Now for $[A\otimes \textbf{x}-\textbf{x}]_i\leq \alpha$ we have 
$\max_{{j^\ast}\in \texttt{fin}_i}\{x_{{j^\ast}}+A(i,{{j^\ast}})\} \leq x_i+\alpha$.
This inequality is satisfied iff all inequality $x_{{j^\ast}}+A(i,{{j^\ast}})\leq x_i+\alpha$ are true. Hence, the corresponding predicates are all true.\QEDB
\end{itemize}
\end{proposition}
\begin{example}
\label{ex1}
Suppose we have an MPL system \eqref{mpl_ex} and $\mathcal{AP}=\{t_1\leq 5\}$. We consider two time-difference specifications $\lozenge  (t_1\leq 5)$ and $\lozenge\square (t_1\leq 5)$ and a set of initial conditions $\mathcal{X}=\mathbb{R}^2$. By \autoref{alg1} and \autoref{alg2}, we have $P_{mat}=\{(1,2,3,1), (1,2,0,1)\}$ and $P_{time}=\{(1,2,0,1)\}$. Thus, $P=\{p_1,p_2\}$ where  $p_1\equiv(1,2,3,1)$ and $p_2\equiv (1,2,0,1)$. 

The resulting abstract transition is depicted in \autoref{fig1}. All abstract states are initial. The corresponding LTL formulae for the time-difference specifications are $\lozenge p_2$ and $\lozenge\square p_2$. 
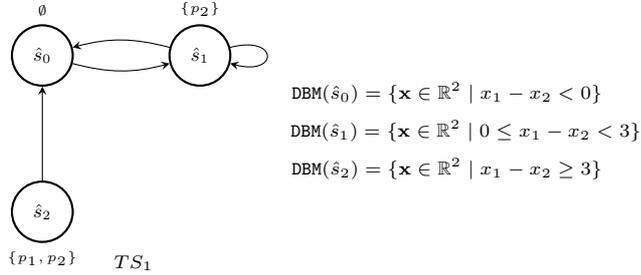
\begin{figure}[!ht]
\centering
\begin{tikzpicture}[node distance=1.25cm and 1.25cm]
\footnotesize
\tikzstyle{place}=[circle,thick,draw=black,minimum size=10mm,scale=0.8]
   \node[place] (1)   {$\hat{s}_0$};
   \node[place ] (2) [right =of 1] {$\hat{s}_1$};
   \node[place] (3) [below =of 1] {$\hat{s}_2$};

   \draw[ ->,>=stealth] (2)  [loop right] to (2);

   \draw[ ->,>=stealth] (3)  to (1);
   \draw[->,>=stealth] (1) [bend right = 15]   to (2);
   \draw[->,>=stealth] (2) [bend right = 15]  to (1);
   \node[above=of 1,yshift=-1.25cm] {\tiny $\emptyset$};
   \node[above=of 2,yshift=-1.25cm] {\tiny $\{ p_2\}$};
   \node[below=of 3,yshift=1.25cm] {\tiny $\{p_1,p_2\}$};
    \node[below=of 3,yshift=1.2cm,xshift=1.2cm] {\scriptsize $TS_1$};
    \node[right of = 2, xshift=2cm,yshift=-0.5cm] {\scriptsize $\texttt{DBM}(\hat{s}_0)= \{\textbf{x}\in \mathbb{R}^2\mid x_1-x_2<0\}$};
    \node[right of = 2, xshift=2.25cm,yshift=-1cm] {\scriptsize $\texttt{DBM}(\hat{s}_1)= \{\textbf{x}\in \mathbb{R}^2\mid 0\leq x_1-x_2<3\}$};
    \node[right of = 2, xshift=2cm,yshift=-1.5cm] {\scriptsize $\texttt{DBM}(\hat{s}_2)= \{\textbf{x}\in \mathbb{R}^2\mid  x_1-x_2\geq 3 \}$};
\end{tikzpicture}
\captionsetup{format=hang,width=0.8\linewidth}
\caption{The abstract transition system via predicate abstractions with a time-difference proposition.}
\label{fig1}
\end{figure}

It is clear that $TS_1\models \lozenge p_2$. Therefore, the underlying MPL system satisfies $\lozenge (t_1\leq 5)$. However, $TS_1\not\models \lozenge \square p_2$ and we can not conclude whether $\lozenge\square (t_1\leq 5)$ is false. We will show how to deal with this problem in Section 4. \QEDB
\end{example}

\subsubsection{Direct Verification} 

In some cases, it is possible to check the satisfaction of time-difference specifications directly, namely without generating the abstraction of the MPL system. 
We call a time-difference proposition $t_i\sim \alpha$ is a \textit{contradiction} if there is no $\textbf{x}\in \mathbb{R}^n$ such that $[A\otimes \textbf{x}-\textbf{x}]_i\sim \alpha$. One the other hand, $t_i\sim \alpha$ is a \textit{tautology} if all $\textbf{x}\in \mathbb{R}^n$ satisfy $[A\otimes \textbf{x}-\textbf{x}]_i\sim \alpha$.
\begin{proposition}
\label{prop_tau_con}
\normalfont
Given an MPL \eqref{mpl} with $A(i,i)=\beta\in \mathbb{R}$. 
\begin{itemize}
\item[i.] For $\sim \{>,\geq\}$, $t_i\sim \alpha$ is a tautology if $\beta\sim\alpha$.
\item[ii.] For $\sim \{<,\leq\}$, $t_i\sim \alpha$ is a contradiction if $\alpha<\beta$. 
\end{itemize}
\textbf{Proof.} One could show that the time difference for the $i^\text{th}$ element is never smaller than the corresponding diagonal element. In other words, for all $\textbf{x}\in \mathbb{R}$ we have $[A\otimes \textbf{x}-\textbf{x}]_i\geq A(i,i)$. Hence, $t_i\geq \beta$ is indeed a tautology. 
\begin{itemize}
\item[i.] The condition $\beta\sim \alpha$ implies $t_i\sim \alpha$ is also a tautology.
\item[ii.] Because $t_i\geq \beta$ is a tautology then its negation $t_i<\beta$ is a contradiction. It is clear that in case of $\alpha<\beta$, $t_i\sim \alpha$ is also a contradiction. \QEDB
\end{itemize}
\end{proposition}
The consequence of \autoref{prop_tau_con} is that any time-difference specification defined from a tautology (resp., contradiction) time-difference proposition, 
is guaranteed to be true (resp., false). For instance, from \autoref{ex1}, the specification $(t_1\geq 2)\mathsf{U}(t_2\geq 3)$ is satisfied, while $\lozenge (t_2\leq 2)$ is not. As a second instance of direct verification, 
in the case of irreducible MPL systems, the dissatisfaction of specifications in the form of $\lozenge\square(t_i\sim \alpha)$ is related to the eigenvalue of the corresponding MPL matrix. 
\begin{proposition}
\normalfont
Consider an MPL system characterised by an irreducible matrix $A\in \Rm{n}{n}$ and a time-difference specification $\lozenge\square(t_i\sim \alpha)$. Suppose $\lambda$ is the max-plus eigenvalue of $A$. The following holds: 
\begin{itemize}
\item[i.] For $\sim \{>,\geq\}$, if $\lambda< \alpha$ then $\lozenge\square(t_i\sim \alpha)$ is false.
\item[ii.] For $\sim \{<,\leq\}$, if $\lambda>\alpha$ then  $\lozenge\square(t_i\sim \alpha)$ is false. 
\end{itemize}
\textbf{Proof.} 
\begin{itemize}
\item[i.] We proof by contradiction. Let assume $\lozenge\square(t_i\geq \alpha)$ is true. Thus, there is an $l\geq 0$ such that 
$[A^{\otimes (k+1)}\textbf{x}-A^{\otimes k}\otimes \textbf{x}]_i\geq  \alpha,~~\forall k\geq l.$
On the other hand, by \autoref{trans_con}, there exists $k_0,c$ such that $A^{\otimes(k+c)}=c\lambda \otimes A^{\otimes k}$ for all $k\geq k_0$. For all $k\geq \max\{k_0,l\}$ we have 
\vspace*{-2ex}\\
$$[A^{\otimes (k+c)}\textbf{x}-A^{\otimes k}\otimes \textbf{x}]_i=\sum_{j=1}^c [A^{\otimes (k+j)}\textbf{x}-A^{\otimes (k+j-1)}\otimes \textbf{x}]_i\geq c\alpha$$
\vspace*{-2ex}\\
One could find that the LHS is equal to $c\lambda$. Hence, we have $\lambda\geq \alpha$ which contradicts $\lambda<\alpha$. From the fact that $\lozenge\square(t_i\geq \alpha)$ is false, it is clear that the strict version of the formula is also false.
\item[ii.] Similar proof of part (i).
\end{itemize}
\end{proposition}

\section{Bounded Model Checking of MPL Systems}

In this section, we implement bounded model checking (BMC) algorithm to check the satisfaction of time-difference specifications over MPL system. The basic idea of BMC is to find a  bounded counterexample of a given length $k$. 
If no such counterexample is found, then one increases $k$ by one until a pre-known completeness threshold is reached, 
or until the problem becomes intractable. 
The readers are referred to \cite{Biere1, Biere2,Biere3} for a more detailed description of BMC.

We use NuSMV 2.6.0 \cite{nusmv} via command $\texttt{check\_ltlspec\_bmc\_onepb}$ to apply BMC. It performs non-incremental BMC to find a counterexample with length 
$k$. If no such bug is present then the command is reapplied for length 
$k+1$, otherwise we apply spurious checking (cf. Section 4.1). In case of non-spurious witness, one can conclude that the time-difference specification is false. Otherwise, we refine the transition system (cf. Section 4.2) such that the counterexample is removed and then reapply BMC command for length $k$. This procedure is repeated until we reach a completeness threshold (cf. Section 4.3).

\subsection{Checking Spuriousness of Counterexamples}

There are two types of $k$-length bounded abstract counterexamples $\pi=\hat{s}_0\hat{s}_1\ldots\hat{s}_k$ in BMC: 
either no-loop or lasso-shaped paths. 
The former one can be used to express the violation of invariant properties $\square p$. A lasso-shaped path is $\pi=\hat{s}_0\hat{s}_1\ldots\hat{s}_k$ such that there exists $1\leq l\leq k$ where $s_{l-1}=s_k$ \cite{Biere1, Biere2}. Although it is finite, it can represent an infinite path $\overline{\pi}=(\hat{s}_0\hat{s}_1\hat{s}_{l-1})(\hat{s}_l\ldots\hat{s}_k)^\omega$ where $\hat{s}_{l-1+m}=\hat{s}_{k+m}$ for $m\geq 0$. It can be used to represent the counterexample of LTL formulae with eventuality, such as $\lozenge p$ and $\lozenge \square p$. 

From now, we write a lasso-shaped path as $(\pi_{stem})(\pi_{loop})^{\omega}$, where $\pi_{stem}=\hat{s}_0\ldots\hat{s}_{l-1}$ and $\pi_{loop}=\hat{s}_l\ldots\hat{s}_{k}$. To avoid ambiguity, we consider that the length of a lasso-shaped path is equal to $|\pi_{stem}|+|\pi_{loop}|$.\footnote{Notice a loop-back transition from $\hat{s}_k$ to $\hat{s}_l$ in $\pi_{loop}$.} Furthermore, any no-loop path cannot be expressed as a lasso-shaped one. That is, if $\pi$ is a no-loop path then the states in $\pi$ are all different. 

The spuriousness of no-loop paths can be checked via forward-reachability analysis. In detail, $\pi=\hat{s}_0\hat{s}_1\ldots\hat{s}_k$ is not spurious iff the sequence of DBMs $D_1,\ldots,D_{k+1}$ where $D_1=\texttt{DBM}(\hat{s}_0)$ and $D_{i+1}=\mathsf{Im}(D_{i})\cap \texttt{DBM}(\hat{s}_{i})$ for $1\leq i\leq k$, are not empty. 
Simply put, there exists $\textbf{x}(0)\in \texttt{DBM}(\hat{s}_0)$ such that $\textbf{x}(i+1)=A\otimes \textbf{x}(i)\in \texttt{DBM}(\hat{s}_{i+1})$ for $0\leq i \leq k$. \autoref{alg5} summarises the procedure of spuriousness checking for no-loop paths.  

\begin{algorithm}[!ht]
 \scriptsize
 \hspace*{\algorithmicindent} \textbf{Input:} $\pi=\hat{s}_0\hat{s}_1\ldots\hat{s}_k$, a no-loop path with length of $k$ \\
\hspace*{\algorithmicindent} \textbf{Output:} $b$, a boolean value \Comment{\scriptsize{$b=\texttt{true}$ iff $\pi$ is spurious}}\\
\hspace*{\algorithmicindent} ~~~~~~~~~~~~~~ $D$, a set of DBMs
\caption{\small Spuriousness checking of no-loop paths}\label{alg5}
\begin{algorithmic}[1]
\Procedure{is$\_$spurious}{$\pi$} 
\State $b\gets \texttt{false}$ 
\State $E \gets \texttt{DBM}(\pi[1])$ \Comment{\scriptsize{$\pi[i+1]=\hat{s}_i$ for $0\leq i\leq k$}}
\State $D\gets \{E\}$  \Comment{\scriptsize{$E$ is the first DBM in $D$}}
\State $k\gets |\pi|-1$
\State $i\gets 1$
\While{($i\leq k$ and $b==\texttt{false}$)}
\State $E\gets \mathsf{Im}(E)\cap \texttt{DBM}(\pi[i+1])$  
\If{$E$ is empty}
\State $b\gets \texttt{true}$
\Else 
\State \textbf{add} $E$ to $D$  \Comment{\scriptsize{$E$ is now the $(i+1)^\text{th}$ DBM in $D$}}
\EndIf
\State $i\gets i+1$
\EndWhile
\State \textbf{return} $(b,D)$
\EndProcedure
\end{algorithmic}
\end{algorithm}

The spuriousness checking for lasso-shaped paths is computed via \autoref{alg6}. We use periodicity checking to deal with the infinite suffix $(\pi_{loop})^\omega$. In lines 14-22, we check the spuriousness of $(\pi_{stem})(\pi_{loop})^{it}$ where $\pi_{loop}$ is repeated $it$ times. If it is not spurious then we check the periodicity of the DBM (line 25). We can conclude that $(\pi_{stem})(\pi_{loop})^{\omega}$ is not spurious if the periodicity is found. In case of an irreducible MPL system, by \autoref{trans_con}, the periodicity is no greater than its cyclicity. On the other hand, after 1000 iterations, if the periodicity cannot be found then the algorithm is stopped with an `undecided' result.

\begin{algorithm}[!ht]
 \scriptsize
 \hspace*{\algorithmicindent} \textbf{Input:} $\pi_{stem}=\hat{s}_0\hat{s}_1\ldots\hat{s}_{l-1}$\\
 \hspace*{\algorithmicindent} ~~~~~~~~~~ $\pi_{loop}=\hat{s}_l\hat{s}_1\ldots\hat{s}_{k}$\\
\hspace*{\algorithmicindent} \textbf{Output:} $b$, a boolean value \Comment{ \scriptsize{$b=\texttt{true}$ iff $\pi$ is spurious}}\\
\hspace*{\algorithmicindent} ~~~~~~~~~~~~~~ $D$, a set of DBMs
\caption{\small Spuriousness checking of a lasso-shaped path}\label{alg6}
\begin{algorithmic}[1]
\Procedure{is$\_$spurious}{$\pi_{stem},\pi_{loop}$} 
\State $(b,D)\gets \texttt{is\_spurious}(\pi_{stem})$ 
 \If{$(b==\texttt{true})$}
 \State \textbf{go to} line 35 \Comment{\scriptsize{$\pi_{stem}$ is already spurious}}
 \Else 
 \State $l\gets |\pi_{stem}|+1$ \Comment{the number of states in $\pi_{stem}$}
 \State $E\gets D[l]$ \Comment{\scriptsize{$E$ is the last DBM in $D$}}
 \State $m\gets |\pi_{loop}|$ \Comment{the number of states in $\pi_{loop}$}
 \State $it\gets 0$ \Comment{the number of iterations}
 \State $p\gets \texttt{false}$ \Comment{\scriptsize{boolean value to represent the periodicity}}
 \While{($it\leq 1000~\text{and}~p==\texttt{false}~\text{and}~b==\texttt{false}$)} \Comment{\scriptsize{maximum number of}}
 \State $it\gets it+1$ \CommentX{\scriptsize{iterations is 1000}\hspace*{4.5ex}}
 \State $i\gets 1$  
 \While{$(i\leq m~\text{and}~b==\texttt{false})$}
 \State $E\gets \mathsf{Im}(E)\cap \texttt{DBM}(\pi_{loop}[i])$   
\If{$E$ is empty}
\State $b\gets \texttt{true}$
\Else 
\State \textbf{add} $E$ to $D$  
\EndIf
\State $i\gets i+1$
 \EndWhile
 \State $j,num\gets |D|$ \Comment{\scriptsize{the number of DBMs in $D$, notice}}
 \While{$(j-m>l ~\text{and}~p==\texttt{false}~\text{and}~b==\texttt{false})$}  \CommentX{\scriptsize{that $\texttt{mod}(|D|,m)=l$} \hspace*{15ex}}
 \If{($D[j-m]==E)$} 
 \State $p\gets \texttt{true}$
 \EndIf
 \State $j\gets j-m$
 \EndWhile
 
 \EndWhile
 \If{($it> 1000~\text{and}~p==\texttt{false}~\text{and}~b==\texttt{false}$)}
 \State \textbf{print} `undecided'
 \Else
 \State \textbf{return} $(b,D)$
 \EndIf
 \EndIf
\EndProcedure
\end{algorithmic}
\end{algorithm}

One can see that the spuriousness checking for no-loop paths (\autoref{alg5}) is guaranteed to be complete. However, this is not the case for \autoref{alg6}. In the case of irreducible MPL systems, it is complete due to the fact that the periodicity is related to \autoref{trans_con}. However for reducible MPL systems, it is incomplete as it may provide undecided results.  

\autoref{lemma_ct} relates the spuriousness of an abstract path, either no-loop or lasso-shaped path, with the value of transient and cyclicity of an irreducible matrix.

\begin{lemma}
\normalfont
\label{lemma_ct}
Consider an irreducible $A\in \Rm{n}{n}$ with transient $k_0$ and cyclicity $c$ and the resulting abstract transition system $TS_f=(\hat{S},T_f,I_f,P_{mat}\cup P_{time},L_f)$. 
Suppose that $\pi$ is a path over $TS_f$. Then, 
\begin{itemize}
\item[i.] If $\pi$ is a no-loop path with $|\pi|\geq k_0+c$, then it is spurious.
\item[ii.] If $\pi=(\pi_{stem})(\pi_{loop})^\omega$ with $|\pi_{stem}|+|\pi_{loop}|> k_0+c$, then it is spurious. 
\end{itemize}
\textbf{Proof.}
\begin{itemize}
\item[i.] Let assume $\pi=\hat{s}_0\ldots\hat{s}_{k_0+c}$ is not spurious. Thus, there exists $\textbf{x}(0)\in \texttt{DBM}(\hat{s}_0)$ such that $\textbf{x}(i+1)=A\otimes \textbf{x}(i)=A^{\otimes i+1}\otimes \textbf{x}(0)\in \texttt{DBM}(\hat{s}_{i+1})$ for $0\leq i\leq k_0+c$. By \autoref{trans_con}, we have $A^{\otimes k_0+c}=\lambda c \otimes A^{\otimes k_0}$ which implies $\textbf{x}(k_0+c)=\lambda c\otimes \textbf{x}(k_0)$. One could show that $\textbf{x}(k_0+c)$ and $\textbf{x}(k_0)$ belong to the same DBM.\footnote{Given a non-empty DBM $D$ and $\alpha\in \mathbb{R}$, if $\textbf{x}\in D$ then so is $\alpha\otimes \textbf{x}$.} Consequently $\texttt{DBM}(\hat{s}_{k_0})=\texttt{DBM}(\hat{s}_{k_0+c})$ and then $\hat{s}_{k_0}=\hat{s}_{k_0+c}$. This contradicts the fact that the states in $\pi$ must be all different.
\item[ii.] Likewise, we assume $\pi=(\pi_{stem})(\pi_{loop})^\omega$ where $\pi_{stem}=\hat{s}_0\ldots\hat{s}_{l-1}$ and $\pi_{loop}=\hat{s}_l\ldots\hat{s}_{k}$ is not spurious. Consequently, there exists $\textbf{x}(0)\in \texttt{DBM}(\hat{s}_0)$ such that $\textbf{x}(i+1)=A\otimes \textbf{x}(i)=A^{\otimes i+1}\otimes \textbf{x}(0)\in \texttt{DBM}(\hat{s}_{i+1})$ for $i\geq 0$.
Again by \autoref{trans_con}, we have $\textbf{x}(i+c)=\lambda c\otimes \textbf{x}(i)$ for $i\geq k_0$. This implies, for $i\geq 0$, $\hat{s}_{k_0+i}=\hat{s}_{k_0+c+i}$. Thus $\pi$ can be rewritten as $(\hat{s}_0\ldots \hat{s}_{k_0})(\hat{s}_{k_0+1}\ldots\hat{s}_{k_0+c})^\omega$. Therefore, the maximum length for $\pi_{stem}$ and $\pi_{loop}$ is $k_0$ and $c$, respectively. This contradicts 
$|\pi_{stem}|+|\pi_{loop}|> k_0+c$.\QEDB
\end{itemize} 
\end{lemma}

\subsection{Refinement Procedure}

Provided that the counterexample is spurious, one needs to refine the abstract transition. 
Instead of adding new predicates as in CEGAR \cite{Clarke6}, 
we are inspired by the refinement procedure described in \cite[Sec. 3.3]{Dieky3}: 
for each abstract state $\hat{s}$ with more than one outgoing transitions, it partitions $\texttt{DBM}(\hat{s})$ according to its successors.  


Our approach for the refinement procedure is slightly different. 
We refine the abstract transition based on a spurious counterexample $\pi=\hat{s}_0\ldots\hat{s}_k$ using the concept of lazy abstraction \cite{Henzinger}. 
This starts by finding a pivot state, namely a state in which the spuriousness starts. Then, it splits the pivot state using the procedure in \cite{Dieky3}. 

Notice that, from \autoref{alg5}, the pivot state can be found from the number of DBMs we have in $D$. One could find that $\hat{s}_{|D|-1}$ is a pivot state. On the other hand, from \autoref{alg6}, a pivot state is $\hat{s}_i$ where $i=|D|-1$, if $|D|<|\pi_{stem}|+1$ (the spuriousness is found in $\pi_{stem}$), otherwise $i=|\pi_{stem}|+1+\texttt{mod}(D-|\pi_{stem}|-1,|\pi_{loop}|)$.

With regards to the refined abstract transitions, the labels and affine dynamics for the new abstract states are equal to those of the pivot state. 
Furthermore, the outgoing (resp. ingoing) transitions from (resp. to) new abstract states are determined similarly using one-step reachability. 

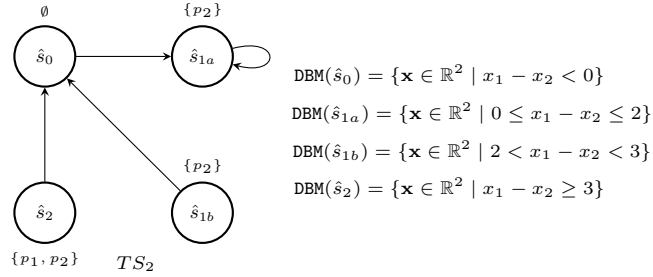
\begin{figure}[!ht]
\centering
\begin{tikzpicture}[node distance=1.25cm and 1.25cm]
\footnotesize
\tikzstyle{place}=[circle,thick,draw=black,minimum size=10mm,scale=0.8]
   \node[place] (1)   {$\hat{s}_0$};
   \node[place ] (2) [right =of 1] {$\hat{s}_{1a}$};
   \node[place] (3) [below =of 1] {$\hat{s}_2$};
   \node[place] (4) [below =of 2] {$\hat{s}_{1b}$};

   \draw[ ->,>=stealth] (2)  [loop right] to (2);

   \draw[ ->,>=stealth] (3)  to (1);
   \draw[->,>=stealth] (4)   to (1);
   \draw[->,>=stealth] (1)     to (2);
   \node[above=of 1,yshift=-1.25cm] {\tiny $\emptyset$};
   \node[above=of 2,yshift=-1.25cm] {\tiny $\{ p_2\}$};
   \node[above=of 4,yshift=-1.25cm] {\tiny $\{ p_2\}$};
   \node[below=of 3,yshift=1.25cm] {\tiny $\{p_1,p_2\}$};
    \node[below=of 3,yshift=1.2cm,xshift=1.2cm] {\scriptsize $TS_2$};
    \node[right of = 2, xshift=2cm,yshift=-0.25cm] {\scriptsize $\texttt{DBM}(\hat{s}_0)= \{\textbf{x}\in \mathbb{R}^2\mid x_1-x_2<0\}$};
    \node[right of = 2, xshift=2.3cm,yshift=-0.75cm] {\scriptsize $\texttt{DBM}(\hat{s}_{1a})= \{\textbf{x}\in \mathbb{R}^2\mid 0\leq x_1-x_2\leq 2\}$};
    \node[right of = 2, xshift=2.3cm,yshift=-1.25cm] {\scriptsize $\texttt{DBM}(\hat{s}_{1b})= \{\textbf{x}\in \mathbb{R}^2\mid  2< x_1-x_2 < 3 \}$};
    \node[right of = 2, xshift=2cm,yshift=-1.75cm] {\scriptsize $\texttt{DBM}(\hat{s}_2)= \{\textbf{x}\in \mathbb{R}^2\mid  x_1-x_2\geq 3 \}$};
\end{tikzpicture}
\captionsetup{format=hang,width=0.75\linewidth}
\caption{The refinement of the abstract transition in \autoref{fig1}. 
The abstract state $\hat{s}_1$ is split into $\hat{s}_{1a}, \hat{s}_{1b}$.}
\label{fig2}
\end{figure}

\begin{example}
We use abstract transition in \autoref{fig1} with specification $\lozenge\square p_2$. The NuSMV model checker reports a counterexample of length 2: $\pi=\hat{s}_1(\hat{s}_0\hat{s}_1)^\omega$. By \autoref{alg6}, it is spurious and the pivot state is $\hat{s}_1$. The resulting post-refinement abstract transition is depicted in \autoref{fig2}. \QEDB
\end{example}

\subsection{Upper-Bound on the Completeness Threshold}

Given a transition system $TS$ and a specification $\varphi$, a completeness threshold is a bound $k$ such that, if no counterexample of $\varphi$ with length 
$k$ or
less can be found in $TS$, then $\varphi$ is satisfied by $TS$ \cite{Biere1,Biere2}. 

We recall from above that for specific formulae, the completeness threshold is related to the structure of the underlying transition system. 
For instance, 
the CT for safety properties of the form $\square p$ is equal to the \textit{diameter} of transition system: the length of longest shortest distance between two states \cite{Biere3}. Likewise, the CT for liveness specifications in the form of $\lozenge p$ is given by the \textit{recurrent diameter} (the length of loop-free path) \cite{Clarke8}. Computing the completeness threshold for general LTL formulae is still an open problem \cite{Clarke8}.


We show that the CT for (abstract) transition system that generated from an irreducible MPL system is related to the transient and cyclicity of the corresponding matrix.

\begin{lemma}
\label{lemma_ct2}
\normalfont
Consider an irreducible $A\in \Rm{n}{n}$ with transient $k_0$ and cyclicity $c$ and the resulting abstract transition system $TS_f=(\hat{S},T_f,I_f,P_{mat}\cup P_{time},L_f)$. 
The CT for $TS_f$ and for any LTL formula $\varphi$ over $P_{mat}\cup P_{time}$ is bounded by $k_0+c$. \QEDB\\
\textbf{Proof.} By \autoref{lemma_ct}, any counterexample of $\varphi$ with length greater than $k_0+c$ (if any) is guaranteed to be spurious.
\end{lemma}

\autoref{lemma_ct2} ensures that the CT is not greater than the sum of the transient and cyclicity of the MPL systems. Looking back to the transition system in \autoref{fig1}, the completeness threshold for $\lozenge p_2$ is 2. In comparison, the transient and cyclicity of matrix in \eqref{mpl_ex} are $k_0=c=2$.

By \autoref{lemma_ct2}, one could say that the BMC algorithm for irreducible MPL systems is complete for any LTL formula. However, this is not the case for reducible MPL systems, due to the incompleteness of \autoref{alg6}.

\section{Computational Benchmarks}

We compare the run-time of the predicate abstractions in this paper with related abstraction procedures in \cite{Mufid}, 
which use max-plus algebraic operations (``tropical abstractions'') and are enhanced versions of the earlier work in \cite{Dieky1}. 
For increasing $n$, we generate matrices $A\in\mathbb{R}_{\max}^{n\times n}$ with two finite elements in each row, each with values ranging between 1 and 10.  
Location and value of the finite elements are chosen randomly. 
The computational benchmark has been implemented on an Intel(R) Xeon(R) CPU E5-1660 v3, 16 cores, 3.0GHz each, and 16GB of RAM.
 
We run the experiments for both procedures using C++.
Over 10 independents experiments for each dimension, 
\autoref{tab1} shows the running time to generate (specification-free) abstractions of MPL systems, 
where entry represents the average and maximal values. 
We do not compare the running time for the generation of abstract transitions because both methods apply the same algorithm. 

\begin{table}[!ht]
\centering
\footnotesize
\captionsetup{format=hang,width=0.75\linewidth}
\caption{Average and maximal running times of abstraction procedures}\label{tab1}
\begin{tabular}{|c|r|r|}
\hline
$n$ & Tropical Abstractions from \cite{Mufid} & Predicate Abstractions (this work)\\ \hline
    3&$\{  0.15,	0.21\}[\text{ms}]$&$\{0.27,	0.38\}[\text{ms}]$\\\hline
4&$\{  0.26,	0.35\}[\text{ms}]$&$\{	0.49,	0.72\}[\text{ms}]$\\\hline
5&$\{0.41,	0.44\}[\text{ms}]$&$\{	0.79,	0.88\}[\text{ms}]$\\\hline
6&$\{  1.12,	1.20\}[\text{ms}]$&$\{	1.92,	2.10\}[\text{ms}]$\\\hline
7&$\{  2.68,	3.74\}[\text{ms}]$&$\{	3.19,	4.60\}[\text{ms}]$\\\hline
 8&$\{  8.78,	10.02\}[\text{ms}]$&$\{9.13,	13.74\}[\text{ms}]$\\\hline
9&$\{  32.12,	36.66\}[\text{ms}]$&$\{30.38,	42.02\}[\text{ms}]$\\\hline
10&$\{  0.12,	0.14\}[\text{sec}]$&$\{0.11,	0.17\}[\text{sec}]$\\\hline
11&$\{  0.57,	0.66\}[\text{sec}]$&$\{0.54,	0.81\}[\text{sec}]$\\\hline
12&$\{  3.82,	4.67\}[\text{sec}]$&$\{2.58,	4.19\}[\text{sec}]$\\\hline
13&$\{  23.71,	28.28\}[\text{sec}]$&$\{	15.80,	28.52\}[\text{sec}]$\\\hline
14&$\{1.39,	1.59\}[\text{min}]$&$\{	0.89,	1.27\}[\text{min}]$\\\hline
15&$\{27.73,	31.06\}[\text{min}]$&$\{	4.68,	8.40\}[\text{min}]$\\\hline
\end{tabular}
\end{table}

As we can see in \autoref{tab1}, for large dimensions (beyond 8), 
the average running time of predicate abstractions is faster than that of tropical abstractions. 
We recall that the (specification-free) predicate abstractions of MPL systems are computed by \autoref{alg1}, \autoref{alg2}, and \autoref{alg4}. 
Whereas for tropical abstractions, they are computed by \cite[Algorithm 2]{Mufid}. 

We also provide a comparison over values of CT. NuSMV is able to compute CT via an incremental BMC command $\texttt{check}\_\texttt{ltlspec}\_\texttt{sbmc}\_\texttt{inc}~\texttt{-c}$. For each bound $k$, in addition to counterexample searching, it generates a SAT (i.e. boolean satisfiability) problem to verify whether the LTL formula can be concluded to hold. This method of computation of completeness check can be found in \cite{Heljanko,Latvala}.

Table 2 shows the comparison of the CT values specified by \autoref{lemma_ct2} and those computed by NuSMV. For dimension of $n\in\{3,4,5\}$, we generate 20 random irreducible matrices $A\in \Rm{n}{n}$ with two finite elements in each row. We use the same time-difference specification $\lozenge \square (t_1\leq 10)$ for all experiments. 
 
\begin{table}[!ht]
\centering
\footnotesize
\caption{The comparison of completeness thresholds.}\label{tab2}
\begin{tabular}{|c|c|c|c|c|}
\hline
$n$ & $ \#\mathtt{stf}$& $\#(\mathtt{ct}_1<\mathtt{ct}_2)$& $\#(\mathtt{ct}_1=\mathtt{ct}_2)$& $\#(\mathtt{ct}_1>\mathtt{ct}_2)$\\ \hline
    3&$14 $&$0$&1&13\\\hline
4&$15$&$1$&0&14\\\hline
5&$14$&$0$&0&14\\\hline
\end{tabular}
\end{table}

The $2^\text{nd}$ column of \autoref{tab2} represents the number of experiments whose the specification $\lozenge \square (t_1\leq 10)$ is satisfied. The last three columns describe the comparison of CT. We use $\mathtt{ct}_1$ and $\mathtt{ct}_2$ to respectively denote the CT that computed by NuSMV and \autoref{lemma_ct2}. As we can see, the CT upper bounds specified by \autoref{lemma_ct2} are relatively smaller than those computed by NuSMV. 

\section{Conclusions} 

This paper has introduced a new technique to generate the abstractions of MPL systems via a set of predicates. The predicates are chosen automatically from system matrix and the time-difference specifications under consideration. Having obtained the abstract states and transition, this paper has implemented bounded model checking to check the satisfaction of time-difference specifications.

The abstraction performance has been tested on a numerical benchmark, which has displayed an improvement over existing procedures. The comparison for completeness thresholds suggests that the cyclicity and transient of MPL systems can be used as an upper bound. Yet, this bound is relatively smaller than the CT bounds computed by NuSMV.  

\subsubsection*{Acknowledgements} 
The first author is supported by Indonesia Endowment Fund for Education (LPDP), while the third acknowledges the support of the Alan Turing Institute, London, UK.  

\bibliographystyle{splncs04}

\bibliography{References}

\end{document}